# Effect of $Pb^{2+}$ substitution on the quantum paraelectric behaviour of $CaTiO_3$

Amreesh Chandra and Dhananjai Pandey

School of Materials Science and Technology, Institute of Technology, Banaras Hindu University, Varanasis-221005, India

## Abstract

Results of low temperature dielectric measurements on the mixed $(Ca_{1-x}Pb_x)TiO_3$ system are presented to show that a dielectric anomaly appears for $x \geq 0.10$. The observed dielectric peak is shown to be due to a thermodynamic phase transition and not due to a relaxor or dipole glass type transition. Curie-Weiss fit to the dielectric constant gives negative Curie temperature which points towards an antiferroelectric transition. Arguments are advanced to propose that $CaTiO_3$ may be an incipient antiferroelectric in which $Pb^{2+}$ substitution stabilises the antiferroelectric phase.

**1. Introduction**

SrTiO$_3$ and KTaO$_3$ are well known incipient ferroelectrics [1-2]. Their dielectric constant gradually increases from 305 and 239 at 300K to very large values of 20,000 and 4500, respectively, upon cooling to 4K [3-5]. The enormous rise in the dielectric constant correlates very well with the experimentally observed softening of the TO mode at q=0 [6-8]. However, below 4K, the dielectric constant in both the materials is independent of temperature. This saturation of dielectric constant below 4K has been attributed to the zero point fluctuations which preclude the condensation of the soft TO mode at q=0. Accordingly, both these materials are said to be quantum paraelectrics below 4K [3]. The possibility of a sharp transition around 40K to a quantum coherent state has also been proposed for SrTiO$_3$ [9-12]. There have been several theoretical efforts to understand these quantum effects in SrTiO$_3$ [3, 13-15]. For example, it has been shown theoretically that quantum zero-point motion can suppress the phase transitions [14]. Using path-integral Monte Carlo simulations and *ab initio* effective Hamiltonian, Zhong and Vanderbilt [15] have recently shown that the quantum fluctuations can indeed suppress the ferroelectric transition of SrTiO$_3$ completely.

Chemical substitutions like A (= Ca$^{2+}$, Ba$^{2+}$ or Pb$^{2+}$) in place of Sr$^{2+}$ in the mixed system (Sr$_{1-x}$A$_x$)TiO$_3$ have been shown to suppress quantum fluctuations and stabilise a quantum ferroelectric phase with T$_c$ varying as (x-x$_c$)$^{1/2}$, where x$_c$≈0.002 [16-17]. In the Sr$_{1-x}$Ca$_x$TiO$_3$ system, the dielectric response gets smeared out for 0.016 ≤ x ≤ 0.12 which has recently been attributed to competing ferroelectric and antiferroelectric instabilities [18-19]. Further, for 0.12≤ x ≤ 0.40, this system exhibits an antiferroelectric phase transition (AFE) [18, 20-21]. Substitutions in place of $K^+$ in the mixed system $(K_{1-x}A_x)TaO_3$ (where A= $Li^{1+}$ or $Na^{1+}$) are also known to suppress the quantum fluctuations but, unlike $(Sr_{1-x}Ca_x)TiO_3$, here a glassy polar phase is stabilised for $x_c \geq$ 0.01 and 0.12 for $Li^{1+}$ and $Na^{1+}$, respectively [1-2]. On increasing their concentration beyond x = 0.022 and 0.20 for $Li^{1+}$ and $Na^{1+}$ substitutions, normal ferroelectric behaviour has been reported [1-2].

A more recent entrant to the family of quantum paraelectrics is $CaTiO_3$ where the saturation of dielectric constant below 30K has been reported [22-23]. However, unlike $SrTiO_3$ and $KTaO_3$, where the role of chemical substitutions in suppressing the quantum fluctuations has been investigated in great detail, there is only one report on the role of chemical substitutions in $CaTiO_3$. Lemanov et al [23] studied the effect of $Pb^{2+}$ and $Ba^{2+}$ substitutions in place of $Ca^{2+}$ in the mixed system $(Ca_{1-x}A_x)TiO_3$ (A=$Pb^{2+}$, $Ba^{2+}$) for x=0.05 and found that the quantum paraelectric behaviour still persists. We present here the results of dielectric measurements on the mixed $(Ca_{1-x}Pb_x)TiO_3$ system for x = 0.10 and 0.20 which show for the first time that a dielectric anomaly corresponding to a thermodynamic phase transition appears for $x \geq 0.10$. It is shown that unlike the $(Sr_{1-x}A_x)TiO_3$ and $(K_{1-x}A_x)TaO_3$ systems, where doping stabilizes polar phases, the dielectric anomaly in the $(Ca_{1-x}Pb_x)TiO_3$ system is not linked with a polar phase (i.e., ferroelectric or dipole glass/ relaxor ferroelectric) transition but may probably be due to an antiferroelectric (antipolar) phase transition(AFE).

**2. Experiment**

Dense (density: 94% of the theoretical density) sintered ceramic pellets of $(Ca_{1-x}Pb_x)TiO_3$ of ≈ 1cm diameter were prepared by the conventional solid state route, the details of which are given elsewhere [24]. Dielectric measurements were carried out on such dense pellets. Both the faces of sintered pellets were gently polished with 0.25μm diamond paste and then washed with acetone to clean off the surface. Isopropyl alcohol was then applied to remove the moisture, if any, left on the pellet surface. Fired-on silver paste was subsequently applied on both the faces of the pellet. It was first dried at 100°C in an oven and then cured by firing at 500°C for 10 minutes. For dielectric measurements below room temperature, a locally designed set-up was used. Temperature was measured with an accuracy of 0.1K using a Kiethley thermometer (model 740). A Hioki LCR meter (Model No. 3532) was used for the measurement of capacitance and loss tangent (tanδ). For collecting room temperature X-ray diffraction (XRD) data, sintered pellets were first crushed into fine powders. These crushed powders were then annealed at 500°C for 10 hours to remove strains which may develop during the process of crushing. These annealed powders were used for collecting the X-ray diffraction (XRD) data. For collecting the XRD data, a 12kW Rigaku make copper rotating anode based powder diffractometer fitted with a curved crystal graphite monochromator in the diffracted beam was used. The data collection was done at 6kW at a scan rate of 1° min$^{-1}$ and step width of 0.01 degree in the 2θ range from 20-120°. The polarization hysteresis loop measurement was done using a locally fabricated Hysteresis Loop Tracer based on a modified Sawer-Tower circuit.

### 3. Results and Discussion

Fig. 1 depicts the room temperature X-ray diffraction patterns of $(Ca_{1-x}Pb_x)TiO_3$ for x = 0.0, 0.10 and 0.20. The XRD patterns in Fig. 1 contain perovskite as well as superlattice reflections marked as P and S, respectively. Both type of reflections can be indexed with respect to a doubled perovskite pseudo-cell. With respect to such a cell, the elementary perovskite peaks assume *hkl* indices which are represented by all even (e) integers. The superlattice reflections, on the otherhand, have one or more odd(o) numbered indices. The patterns shown in Fig. 1 contain superlattice reflections of the type *ooo, ooe and oee* (e.g. 311, 312, 322 in the figure) which arise due to antiphase octahedral tilts, inphase octahedral tilts and antiparallel cationic ($Ca^{2+}$/ $Pb^{2+}$) displacements, respectively (see Reference [25]). Since all the perovskite and superlattice peaks are common for x=0, 0.10 and 0.20, we conclude that there is no change in the room temperature structure of $CaTiO_3$ as a result of $Pb^{2+}$ substitution i.e, the crystal structure remains orthorhombic with Pbnm space group and $a^-a^-c^+$ tilt system [25]. The cell parameters as obtained by least squares refinement are a = 5.39(7) , 5.43(1) and 5.45(1) Å, b = 5.410(7), 5.41(1) and 5.45(1) Å, and c = 7.65(1), 7.68(1) and 7.70(4) Å for x=0.0, 0.10 and 0.20 respectively. The unit cell volume increases with x as expected on the basis of larger ionic radius of $Pb^{2+}$ as compared to that of $Ca^{2+}$.

The dielectric constant of $CaTiO_3$ increases with decreasing temperature as per Barret law[26] and shows quantum saturation below 30K as expected for a quantum paraelectric [22]. The effect of $Pb^{2+}$ doping on the temperature dependence of dielectric constant of $CaTiO_3$ is shown in Fig. 2(a and b) which depicts the temperature variation of real ($\varepsilon'$) and imaginary ($\varepsilon''$) parts of the dielectric constant for $(Ca_{1-x}Pb_x)TiO_3$ with x=0.10 and 0.20. It is evident from this figure that a dielectric peak appears at 147K and 164K for x=0.10 and 0.20, respectively. It is interesting to note that the dielectric peak in Fig. 2(a and b) appears on a continuously rising background due to the paraelectric behaviour of the $CaTiO_3$ matrix. The height of the dielectric peak increases with increasing $Pb^{2+}$ content. This dielectric anomaly is not due to a relaxor/dipole glass type transition since the temperatures $T_m'$ and $T_m''$ corresponding to the peak values of $\varepsilon'$ and $\varepsilon''$ are coincident (see Fig. 2). In relaxor/ glassy systems, it is well known that $T_m'' < T_m'$ and both of them shift to higher temperature sides on increasing the frequency of measurement [27,28]. In the $(Ca_{1-x}Pb_x)TiO_3$ system, $T_m''$ and $T_m'$ are not only coincident but also frequency independent as can be seen from Fig. 3 which depicts $\varepsilon'$ and $\varepsilon''$ at various frequencies upto 200kHz. The small frequency dispersion (~5%) in the dielectric constant at all temperatures in the frequency range 10-200kHz is due to conductivity losses. These losses arise due to $Pb^{2+}$ vacancies which are unavoidable at the sintering temperature (1473 K) as a result of high vapour pressure of PbO above 1073K. This dispersion is not due to relaxor ferroelectric behaviour for which it should have been confined to the vicinity of the transition temperature[28]. The frequency independence of $T_m'$ and $T_m''$ further rules out any relaxor/ dipolar glass behaviour. Thus the dielectric anomaly shown in Fig. 2 is due to a thermodynamic phase transition and not due to relaxor/glassy transition.

In order to check if the observed dielectric anomaly in Fig. 2(a,b) is due to a ferroelectric transition, we carried out polarization hysteresis loop measurements above and below the transition temperatures on the $(Ca_{1-x}Pb_x)TiO_3$ samples. No hysteresis loop was observed between polarization and electric field below the transition temperature for fields upto 60kV/cm. Dielectric breakdown of the sample occurred on application of fields greater than 60kV/cm. For conventional ferroelectrics and dipolar glass/ relaxor ferroelectric materials, an external field of 60kV/cm is large enough to open the hysteresis loop[2,27,28]. The absence of P-E hysteresis loop below the transition temperature clearly rules out the possibility of a ferroelectric transition being responsible for the observed dielectric anomalies in Fig. 2.

The dielectric constant of $(Ca_{1-x}Pb_x)TiO_3$ system obeys Curie-Weiss behaviour at high temperatures as shown in Fig. 4 (a and b) for x = 0.10 and 0.20 respectively. The departure from Curie-Weiss law over about 20K range (shown by dotted vertical lines in Fig. 4) above the transition temperature points towards diffuse nature of this transition which may be due to local compositional fluctuations which are invariably present in samples prepared by conventional solid state route [27]. Since the transition temperature ($T_m^{/}$) of the mixed system varies with Pb content (x) at a rate of 3.3K/mol% of $PbTiO_3$, different regions of the sample with slightly varying compositions will have different transition temperatures ($T_m^{/}$). The experimentally measured $\varepsilon^{/}$ versus temperature plot will therefore be a broad envelope of these local transitions. This will naturally smear out the $\varepsilon^{/}$ versus temperature plots. Such diffuse transitions are known to exhibit departure from Curie-Weiss behaviour over wide range of temperatures [29,30]. The Curie-Weiss temperatures ($T_c$) as obtained from the extrapolation of the linear region in the $1/\varepsilon^{/}$ versus temperature plots in Fig. 4 are found to be –109K and –15K for x=0.10 and 0.20, respectively. The significance of negative Curie-Weiss temperature in this mixed system needs to be elaborated.

For ferroelectric materials, the dielectric constant above the transition temperature ($T_m^{/}$) is known to follow Curie-Weiss behaviour:

$$\chi^{/} = C / (T-T_c) \quad \text{--------------------(1)}$$

where C is the Curie constant and $T_c$ the Curie Weiss temperature. For ferroelectrics, $T_c$ has to be necessarily positive and it may be less than or equal to $T_m^{/}$ for first and second order phase transitions, respectively [31]. For antiferroelectrics, phenomenologically, the dielectric constant follows the relationship [32]

$$\varepsilon' = 1/[g + \lambda(T - T_0)] \text{ ---------------- (2)}$$

where g is a measure of the coupling between the sublattice polarization (antipolar), $\lambda$ is inverse of Curie constant. Eq. (2) can be recast in the form of Eq. (1) by putting $1/\lambda = C$ (C is Curie constant) and $T_0 - g/\lambda = T_c$. Unlike ferroelectrics where $T_o$ is necessarily positive, the sign of $T_c$ for antiferroelectrics may be positive or negative depending on whether $T_0<gC$ or $T_0>gC$, respectively i.e., it depends on the value of the Curie-Weiss constant (C) and the sublattice antipolar coupling(g) [21]. The negative value of Cure-Weiss temperature for $(Ca_{1-x}Pb_x)TiO_3$ implies that the dielectric anomaly in Fig. 2 is definitely not due to a ferroelectric transition but can be due to an antiferroelectric phase transition. This is similar to the case of $(Ca_{1-x}Sr_x)TiO_3$ system where for compositions $0.88 < x \leq 0.40$, negative value of $T_c$ has been shown to be due to an antiferroelectric transition [18, 20-21]. In magnetic systems also, negative value of $T_c$ is taken as a proof for an antiferromagnetic transition[33].

At this stage, it may be pertinent to understand as to why $(Ca_{1-x}Sr_x)TiO_3$ [18,20-21] and $(Ca_{1-x}Pb_x)TiO_3$ systems exhibit antiferroelectric transition with negative $T_c$. As far as one of the end members ($SrTiO_3$ or $PbTiO_3$) is concerned, it is either an incipient ferroelectric or ferroelectric but does not possess any antiferroelectric instability. In the following, we shall now advance arguments to propose that the origin of antiferroelectricity in these two mixed systems lies in the incipient antiferroelectric behaviour of $CaTiO_3$. The temperature variation of the dielectric constant in $SrTiO_3$ [3],

KTaO$_3$[8] and CaTiO$_3$ [22] have been fitted with the following quantum mechanical mean-field formula due to Barrett [26]

$$\chi = M / [(T_1/2) \coth (T_1/2T) – T_c] \text{ ------------- (3)}$$

This expression has been derived within the self consistent single mode theory for quantum crystals [3] as well as a limiting case of a renormalized harmonic approximation involving nonlinear polarizability of the oxygen shell [4]. The characteristic temperature $T_1$ in Eq. (3) signifies the onset of low temperature region in which the quantum effects are important and dielectric constant shows deviation from the classical Curie-Weiss law. The values of $T_1$ = 84K and 56.9K, and $T_c$= 38K and 13.1K for SrTiO$_3$ and KTaO$_3$, respectively, are positive [3-4] whereas for CaTiO$_3$ $T_1$=104K and $T_c$= -159K [16]. Dielectric measurements by other workers [23,34] have also revealed negative $T_c$ for CaTiO$_3$. For temperatures T>$T_1$ where the quantum effects can be ignored, Eq.(3) reduces to the classical Curie-Weiss law with $T_c$ of Eq.(3) becoming the Curie-Weiss temperature[26]. A negative $T_c$ for CaTiO$_3$ clearly rules out incipient ferroelectric behaviour proposed by earlier workers [23,35] but implies an incipient antiferroelectric behaviour. This is in contrast to SrTiO$_3$ and KTaO$_3$ which are incipient ferroelectrics with positive $T_c$. In a recent experimental study of the temperature dependence of soft mode frequency in CaTiO$_3$, Zelezny et al [36] have also reported negative $T_c$ of about –105K which they have attributed to an incipient ferroelectric behaviour. The arguments advanced in this paper, however, clearly show that $T_c$ has to be necessarily positive for incipient or regular ferroelectric behaviour. We suspect that the soft mode observed by these workers corresponds to a non-zone centre phonon which is responsible for the negative $T_c$. It seems that Pb$^{2+}$ and Sr$^{2+}$ substitutions stabilize the AFE phase in the CaTiO$_3$ matrix by raising the transition temperatures to values greater than T$_1$(=104K) in Eq.(3) such that the effect of quantum fluctuations becomes negligible. The present results also suggest that the recent first principles calculations [37-38] of phonons and static dielectric constant in CaTiO$_3$ need to be reexamined from the point of view of an antiferroelectric instability.

## 4. Conclusions

The dielectric constant of CaTiO$_3$ increases with increasing temperature as per Barret's law with evidence for quantum saturation below 30K. We have shown for the first time that Pb$^{2+}$ substitution can lead to a dielectric anomaly for concentrations x≥0.10. Further, we have shown that the dielectric anomaly is due to a thermodynamic phase transition and not due to a relaxor/ dipolar glass transition. The Curie-Weiss fit to the dielectric constant above the transition temperature reveals negative Curie-temperature which implies an antiferroelectric transition. It is argued that CaTiO$_3$ is an incipient antiferroelectric in which Pb$^{2+}$ substitution suppresses the quantum fluctuations and thereby stabilises the antiferroelectric phase.

## 5. Acknowledgement

We thank IUC-DAEF for partial financial support.

**FIGURE CAPTIONS.**

Fig.1 XRD patterns of $(Ca_{1-x}Pb_x)TiO_3$ with x = (a) 0.20, (b) 0.10 and (c) 0.00. The y-axis is sufficiently zoomed to enable the superlattice peaks to become visible. Because of this, some of the main perovskite peaks appear truncated.

Fig. 2 Variation of real ($\varepsilon'$) and imaginary ($\varepsilon''$) parts of the dielectric constant with temperature at 10 kHz in $(Ca_{1-x}Pb_x)TiO_3$ ceramic samples with (a) x=0.10 and (b) x=0.20

Fig. 3 Variation of dielectric constant with temperature at frequencies 10, 50 and 200 kHz in $(Ca_{1-x}Pb_x)TiO_3$ ceramic sample with (a) x=0.10 and (b) x=0.20. The filled and open symbols, respectively, represent the variation of real($\varepsilon'$) and imaginary($\varepsilon''$) parts of dielectric constant .

Fig. 4 Curie-Weiss fit to $\varepsilon'(T)$ data for $(Ca_{1-x}Pb_x)TiO_3$ with (a) x=0.10 and (b) x=0.20. Vertical dashed lines show the departure region near the transition temperature.

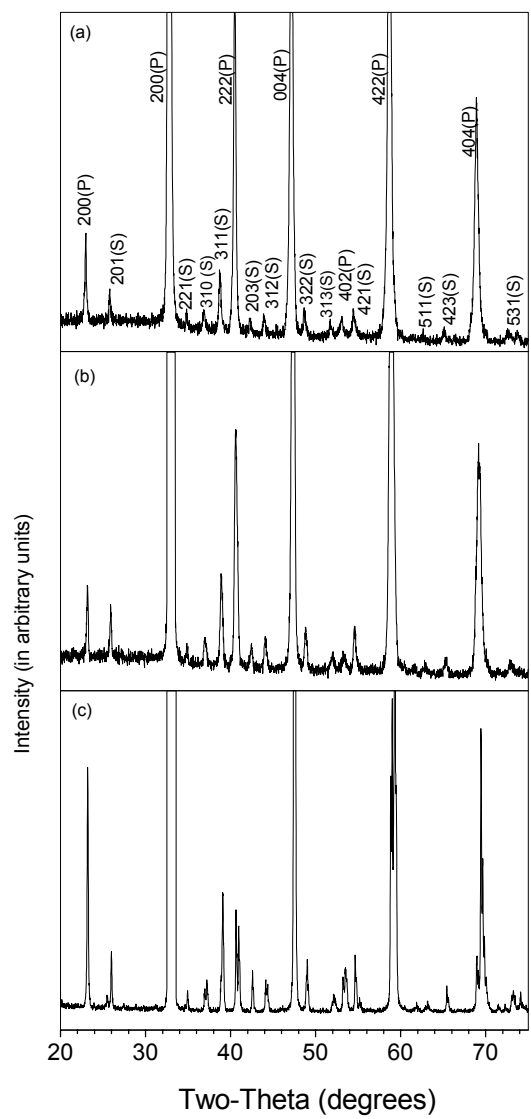

Fig.1

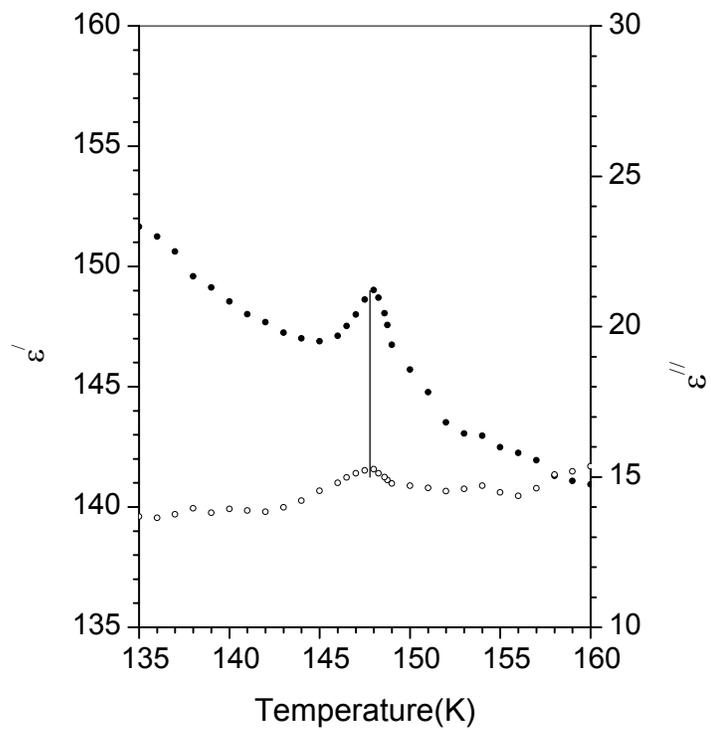

Fig.2(a)

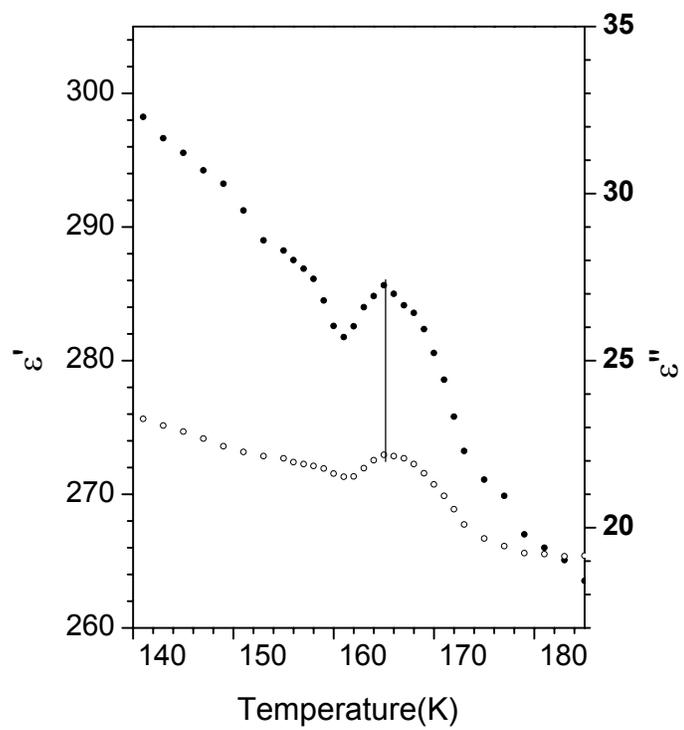

Fig. 2(b)

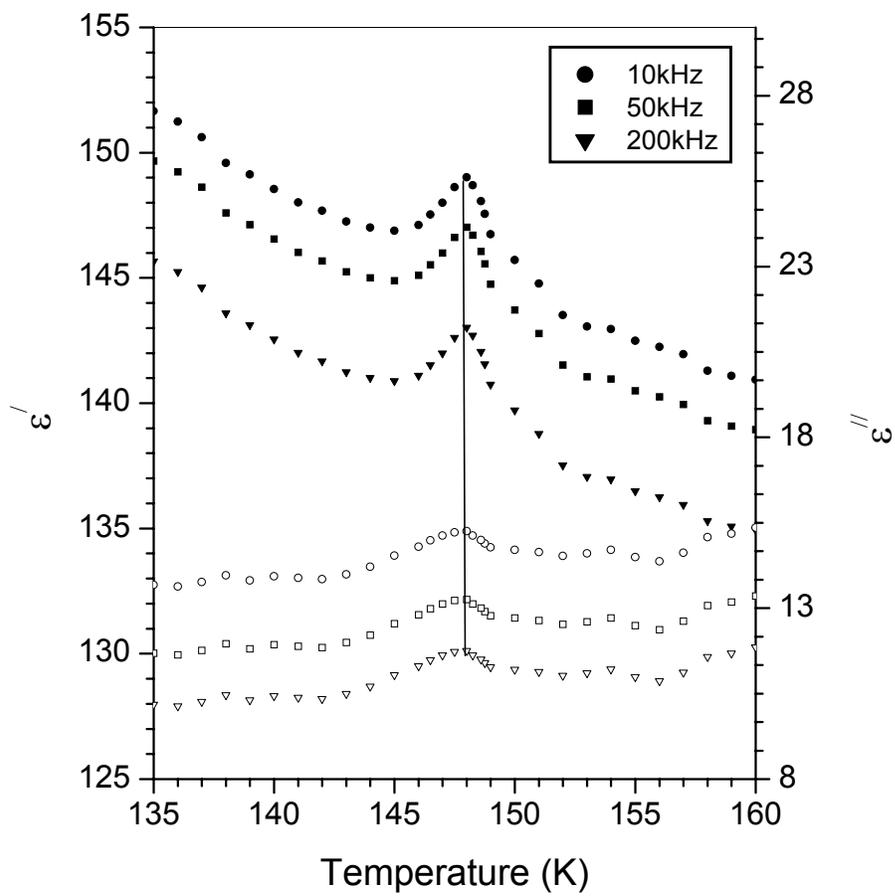

Fig.3(a)

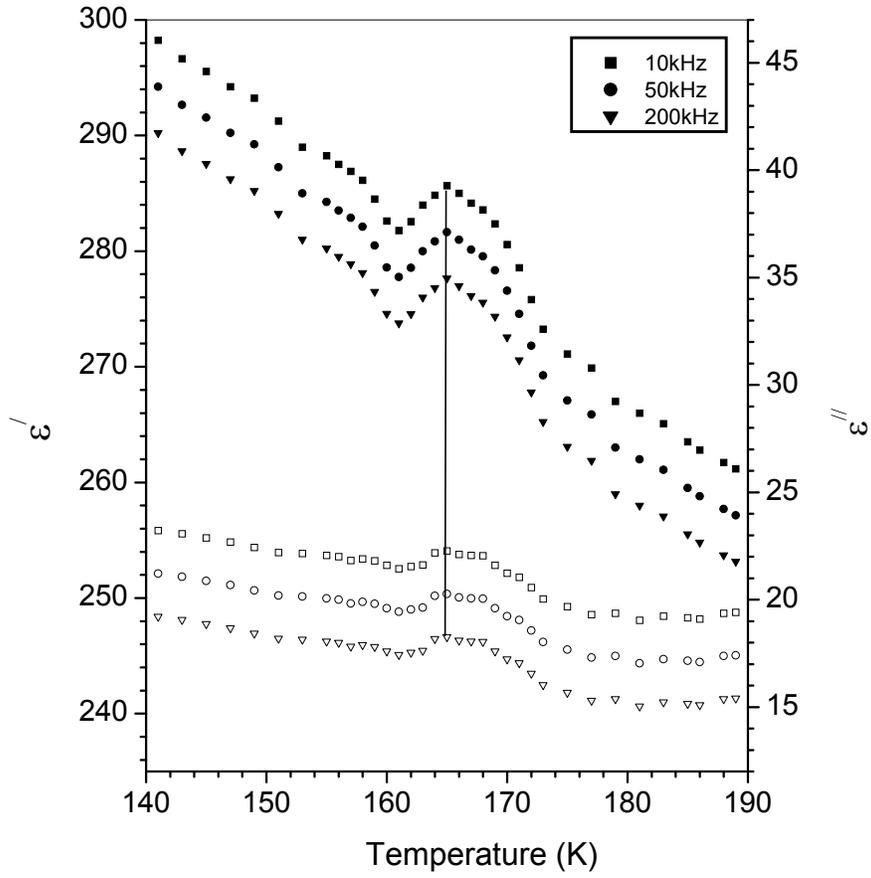

Fig.3(b)

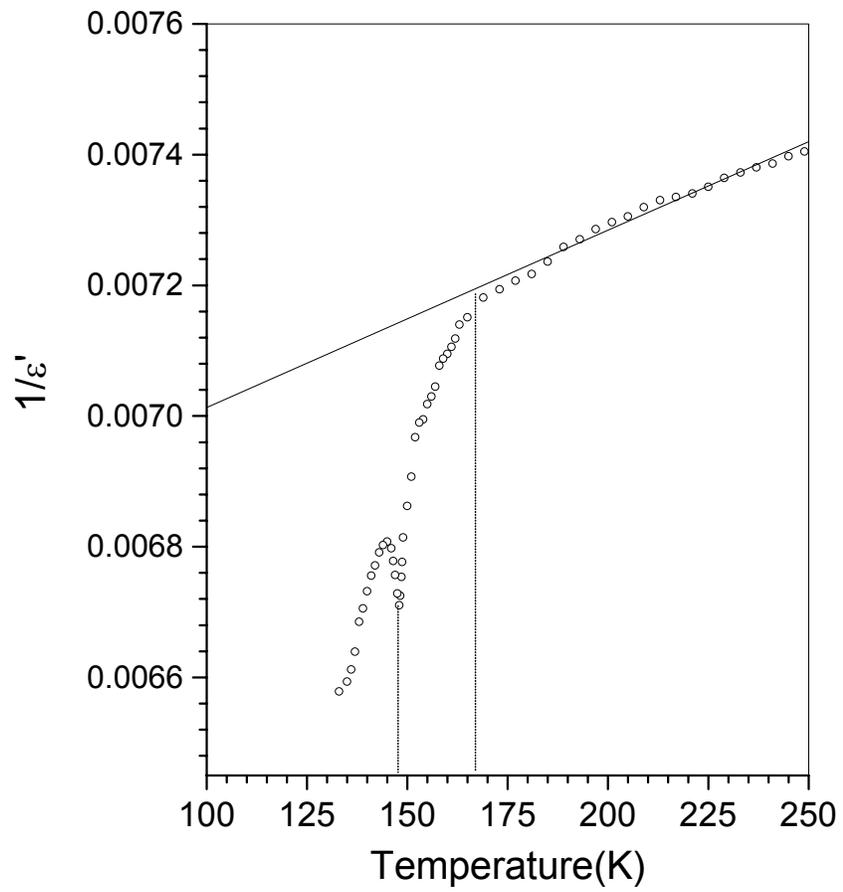

Fig.4(a)

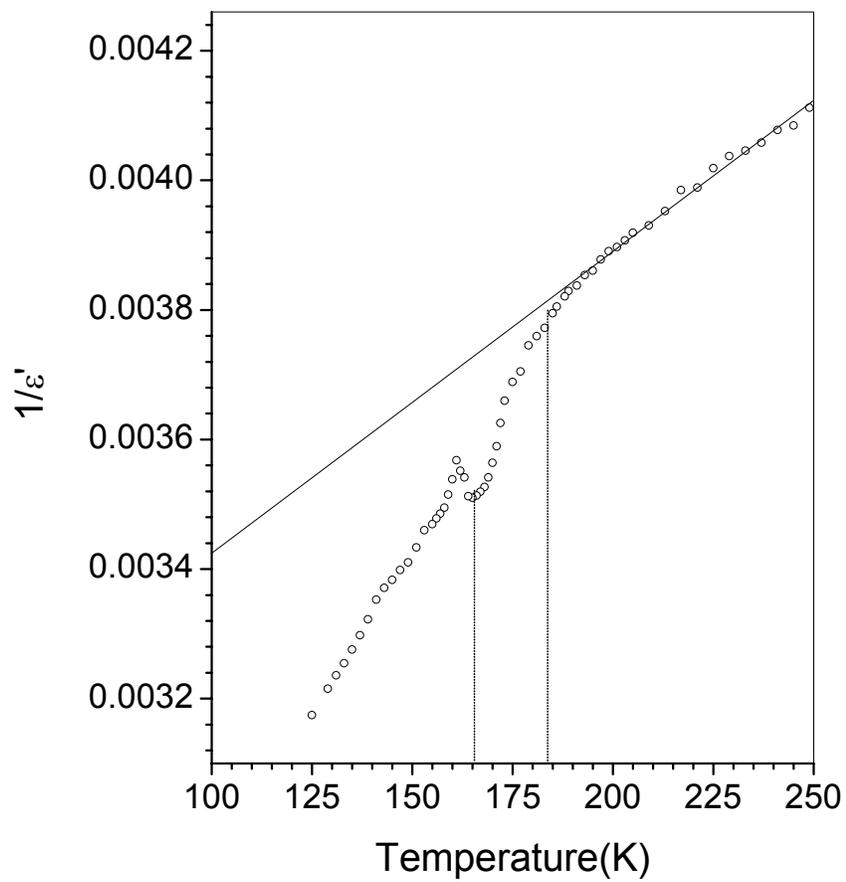

Fig. 4(b)